\documentclass[pdflatex,sn-mathphys,iicol]{sn-jnl}% Math and Physical Sciences Reference Style
\usepackage{subfigure}
\usepackage{graphics}
\usepackage{xcolor}
\usepackage{amsmath}
\jyear{2022}%

\theoremstyle{thmstyleone}%
\theoremstyle{thmstyletwo}%

\theoremstyle{thmstylethree}%

\begin{document}

%\title[Investigation of spin rotators in CEPC at Z-pole]{Investigation of spin rotators in CEPC at Z-pole}

\title[*]{Investigation of spin rotators in CEPC at the Z-pole}


\author[1,2]{\fnm{Wenhao} \sur{Xia}}

\author*[1]{\fnm{Zhe} \sur{Duan}}\email{duanz@ihep.ac.cn}

\author*[1,2]{\fnm{Jie} \sur{Gao}}\email{gaoj@ihep.ac.cn}
\author*[1]{\fnm{Yiwei} \sur{Wang}}\email{wangyw@ihep.ac.cn}

\affil[1]{\orgdiv{Key Laboratory of Particle Acceleration Physics and Technology}, \orgname{Institute of High Energy Physics,Chinese Academy of Sciences}, \orgaddress{\street{19B Yuquan Road}, \city{Beijing}, \postcode{100049},  \country{China}}}

\affil[2]{ \orgname{University of Chinese Academy of Sciences}, \orgaddress{\street{19A Yuquan Road}, \city{Beijing}, \postcode{100049},  \country{China}}}


%%================================%%
%% Sample for structured abstract %%
%%================================%%

\abstract{\textbf{Purpose:} Longitudinal polarization is an important design aspect of the future 100 
 km-scale Circular Electron Position Collider (CEPC). Spin rotators are needed in the CEPC collider rings to make the beam polarization along the longitudinal direction at the interaction points (IPs). This paper focuses on the design of spin rotators for CEPC at the Z-pole (45.6~GeV). 

\textbf{Methods:} The design of spin rotators in the CEPC at the Z-pole is based on solenoid magnets and horizontal bending magnets sections. The coupling of transverse motion introduced by solenoids is compensated with quadrupole lenses.
Adjustments have been made to the layout to implement the spin rotators
into the collider rings.

\textbf{Results:}  Longitudinal polarized beam can be achieved at the IPs with the spin rotators. High degree of polarization is attainable, while the effect of spin rotators on orbital motion is acceptable.  The detailed simulation results will be presented.
 
\textbf{Conclusion:} A solenoid-based spin rotator configuration is designed and integrated into the CEPC collider ring lattice. According to the simulation results, the polarization requirements can be satisfied.
}




\maketitle

\section{Introduction}
As a future lepton circular collider, the Circular Electron Positron Collider (CEPC) is proposed for testing the underlying fundamental physics principles
of the Standard Model (SM) and exploring physics beyond the SM~\cite{cepcsnowmass}. In the Conceptual Design Report (CDR)~\cite{cepccdr}, the CEPC is designed as a 100-km double-ring  collider with two interaction points (IPs). The operated beam energies are 120 GeV as a Higgs factory, 80 GeV as a W factory and 45.6 GeV for the Z-pole.

Particle physics experiments with longitudinally polarized beams in the CEPC is an important design aspect for an accurate check of the Standard Model.
High longitudinal beam polarization ($>$50\%) is required in the interaction points of the CEPC without sacrificing luminosity. In this paper,
we focus on the realization of longitudinal polarization at the Z-pole. The possibility at even higher energies will be addressed elsewhere.

To realize longitudinally polarized colliding beams, we need to address two major issues,
polarized beam generation and proper spin rotations in the collider ring.
A highly polarized electron beam ($>80\%$) can be generated by a polarized electron gun~\cite{egun}, accelerated to full collision energy by the Linac and the Booster. There will be polarization decay during the acceleration in the Booster. A preliminary design of Siberian snakes is under investigation to overcome the depolarization in the Booster~\cite{duanz2}. It is an independent research topic and therefore it will be not covered here.
Finally the polarized beams are injected into the collider rings. 
Technically speaking, polarized positron sources are
costly or less matured to be applied to CEPC-Z at the moment~\cite{e+source1,e+source2,duanz2}. 
%{\color{red} And according to experiments on colliding electron-positron beams~\cite{Nikitin:1981rb, lepslc},  it is already very useful to have polarized only one of the counter beams, like longitudinal polarized  electron beams alone.}
And in the vast majority of cases, single beam polarization is sufficient, since  annihilation occurs when the particles are of opposite helicity, like in 
the measurement of polarized lepton asymmetries at SLC~\cite{lepslc}. An electron 
with a certain helicity will find a partner for itself. Therefore, it is 
sufficient to have only one of the colliding beams longitudinally 
polarized, for example, an electron beam.
But there may be technological innovations in the future, which can provide a positron beam with a high degree of polarization and meet the bunch intensity requirements. Then the scheme of electron beam polarization maintenance in the Booster also applies to the polarized positron beam. Hence, we design the longitudinal polarization scheme for both
$e^-$ and $e^+$ beams.
In this paper, spin rotators are designed for $e^+/e^-$ beams in the CEPC collider ring at the Z-pole, to realize the longitudinal polarized beam at the IPs in the collider rings.

A spin rotator is a device which can rotate the particle spins from vertical to longitudinal or vice versa in storage rings. A pair of rotators around each IP are required to keep the polarization vertical in the arcs and to realize longitudinal polarization at the IP. Spin rotators have been successfully employed
at HERA for electron and positron beams~\cite{herarotator}. The design of the HERA spin rotators is known as the `mini rotator'~\cite{minirotator} which consists of a sequence of alternating vertical and horizontal dipoles. In addition, Spin rotator schemes based on helical dipole magnets have been adopted in RHIC to achieve longitudinal polarization at the interaction points~\cite{rhicrotator}. 
However, the beam orbit excursion may be large in those spin rotators with pure transverse magnetic fields.
Hence the magnet bore must be large enough to accommodate the excursions. Besides,  vertical bending magnetic fields in the spin rotators also introduces vertical dispersion and thus an increase in the vertical emittance. This can lead to a reduction of the equilibrium 
beam polarization,
and might not be compatible with high-luminosity $e^+/e^-$
circular collider designs which favor very flat beams.
In addition, a combination of solenoids and horizontal bending magnets 
can also form
a spin rotator, hereafter referred to as ``solenoid-based spin rotators''.

A spin rotator including a dedicated horizontal bending magnet section and a dedicated solenoid section, was an alternative design first studied for HERA~\cite{barber}, where the spin matching conditions were first derived for solenoid-based spin rotators. Later, similar concept was also applied for eRHIC~\cite{erhicbarber}. Koop also discussed this kind of spin rotators for the SuperB project~\cite{koop,superb} later. 
Nikitin first pointed out such a scheme could be modified and applied in the CEPC at the Z-pole energy ~\cite{nikitin2,nikitin3}, using the fact 
that the interaction region design features a S-shaped twisted orbit in 
median plane, i.e., the spin rotators can be arranged in an ``anti-symmetric'' manner, which
cancels most of the depolarization effect due to quantum fluctuation~\cite{barber}.
Recently, a solenoid-based spin rotator for the electron polarization in the electron-ion collider (EIC) is under study~\cite{eiccdr}. The spin rotator scheme for the electron polarization uses two pairs of solenoid magnets and associated dipoles
on each side of the interaction region. And the design of spin rotators
for the electron storage ring allows to realize the exact longitudinal orientation of electron spins in the whole energy range from 5 to 18~GeV,
via adjustment of solenoid strengths without changing the machine geometry~\cite{erhic,jleic}. 
For SuperKEKB, a very compact spin rotator design is under study that consists of solenoid-dipole combined function magnets~\cite{superkekb}.

In order to avoid those problems introduced by transverse magnetic fields, we will study a solenoid-based spin rotator configuration in this paper. Then its integration into the CEPC collider ring is investigated. In this study, we used the CEPC CDR lattice at the Z-pole. The SAD code~\cite{sad} was used in the design of the spin rotator and evaluation of its influence to orbital motion. The Bmad/PTC code~\cite{Bmad,PTC} was used to evaluate the performance of the spin motion and polarization.
It has been verified that lattice translations between SAD and Bmad/PTC provide consistent results by Zhou \textit{et~al}~\cite{zhoudm}.

The theories of spin dynamics are reviewed in Section~\ref{sec2}. Section~\ref{sol} presents the designs of solenoid-based spin rotators. In Section~\ref{opt}, the insertion scheme of the spin rotators is studied, and the simulation results for spin motion and orbital motion are presented. Finally, potential optimization of the spin rotators insertion is analyzed qualitatively.

\section{Theory\label{sec2}}
A beam is polarized if there exists a direction for which the two possible spin states are not equally populated.
Polarization is the statistical average of an  ensemble of the spin of particles in the beam. 
Consider a beam consists if $N$ electrons,
the polarization degree along a specified axis is given by
\begin{equation}
    P=\frac{N_{\uparrow}-N_{\downarrow}}{N}
\end{equation}
where $N_{\uparrow}+N_{\downarrow}=N$. $N_{\uparrow}$ and $N_{\downarrow}$ are the number of electrons in the up state and down state respectively.

The precession of the spin vector $\mathbf{S}$ of a charged particle with the velocity $\mathbf{v}$ and spin-$\frac{1}{2}$ in an electromagnetic field is determined by the Thomas-BMT equation~\cite{thomas,bmt}:
\begin{eqnarray}
   \frac{{\rm d}\mathbf{S}}{{\rm d}t}&=&\mathbf{\Omega} \times \mathbf{S} \nonumber \\
    \mathbf{\Omega}&=&-\frac{e}{m_0\gamma}\left[ (1+a\gamma)\mathbf{B_{\perp}}+(1+a)\mathbf{B_{\parallel}}\right. \nonumber \\
    &+&\left.(a\gamma+\frac{\gamma}{1+\gamma})\frac{\mathbf{E}\times \boldsymbol{\beta}}{c^2}\right ] 
    \label{eq:spinmotion}
\end{eqnarray}
where $\mathbf{B}=\mathbf{B_{\parallel}}+\mathbf{B_{\perp}}$ , $\mathbf{B_{\parallel}}=(\boldsymbol{\beta}\cdot\mathbf{B})\boldsymbol{\beta}/\beta^2$. $\boldsymbol{\beta}=\mathbf{v}/c$. $m_0$ and $e$ are the mass and charge of the particle, respectively. $a$ is the anomalous magnetic moment, $a=0.00115965$ for electron and positron. $\gamma$ is the relativistic factor. $c$ is the speed of light. 

When $t$ is replaced by the azimuthal angle $\theta$, one special solution of Eq.(\ref{eq:spinmotion}) is the invariant spin field \cite{barber2,barber3}, $\mathbf{n}(\mathbf{u};\theta)$, satisfying the periodic condition $\mathbf{n}(\mathbf{u};\theta+2\pi)=\mathbf{n}(\mathbf{u};\theta)$, $\mathbf{u}$ is the phase space coordinate in terms of
canonical variables of the orbital motion.
The rate of spin precession around $\mathbf{n}$ is described by
the amplitude dependent spin tune $\nu_s$ \cite{barber2,barber3}. On the closed orbit, $\mathbf{n}(\mathbf{u};\theta)$ and $\nu_s$ reduces to $\mathbf{n}_{0}(\theta)$ and $\nu_0$, respectively. In a planar storage ring, 
 $\mathbf{n}_{0}(\theta)$ is close to the vertical direction. And $\nu_0 \approx a \gamma_0$ where $\gamma_0$ is the relativistic factor for the design energy.

The spin motion may be perturbed in a resonant manner when the following condition is satisfied,
\begin{equation}
    \nu_s=k+k_x\nu_x+k_y\nu_y+k_z\nu_z,~~~k,k_x,k_y,k_z \in Z
\end{equation}
where $\nu_x, \nu_y, \nu_z$ are the horizontal, vertical betatron tunes and synchrotron tune, respectively. Spin resonances with $\nu_s=k, k \in Z$ are called the integer spin resonances. Spin resonances with $\lvert k_x\rvert+\lvert k_y\rvert+\lvert k_z\rvert=1$
and $\lvert k_x\rvert+\lvert k_y\rvert+\lvert k_z\rvert > 1$ are called the first-order spin resonances and the higher-order spin resonances, respectively. In particular, synchrotron sideband spin resonances $\nu_0\pm\nu_u+m\nu_z=k$, with $\mid m \mid>1$ and $u=x,y,z$ are considered very important in high-energy electron storage rings~\cite{mane}.

In an electron storage ring, a spontaneous buildup of radiative polarization may occur via the emission of spin-flip synchrotron radiation. This process is called the Sokolov-Ternov effect~\cite{st}. However, the stochastic photon emission also
leads to the spin diffusion effect~\cite{bks}.
The equilibrium polarization ~\cite{dk} is a balance between the Sokolov-Ternov effect and the spin diffusion effect, its direction is along
$\left<\mathbf{n}\right>_{\theta}$, which is an average over phase space at the azimuthal angle ${\theta}$. Generally speaking, $\left<\mathbf{n}\right>_{\theta}$ is very
nearly aligned along $\mathbf{n}_{0}(\theta)$.
The degree of the equilibrium polarization can be approximated by
\begin{equation}
P_{\rm eq}\approx \frac{P_{\infty}}{1+\frac{\tau_{\rm p}}{\tau_{\rm d}}}
\label{eq:Peq}
\end{equation}
where $P_{\infty}$ equals to 92.4\% in an ideal planar ring.  And the equivalent polarization build-up time is
\begin{equation}
\frac{1}{\tau_{\rm tot}}=\frac{1}{\tau_{\rm p}}+\frac{1}{\tau_{\rm d}}
\label{eq:tau_tot}
\end{equation}
$\tau_{\rm p}$ and $\tau_{\rm d}$ are the time constants for the Sokolov-Ternov effect and spin diffusion effect, respectively. 
For the CEPC collider ring at the Z-pole. $\tau_p\approx 260$ hours. 
According to Eq.(\ref{eq:Peq}) and (\ref{eq:tau_tot}), if the equilibrium polarization equals to 50\%, we can get $\tau_d\approx 310$ hours and the equivalent polarization build-up time $\tau_{tot}\approx141$ hours.

For the CEPC collider rings, top-up injection mode will be used to maximize the integrated luminosity~\cite{cepccdr}. Time-averaged beam polarization in an electron storage ring during the top-up injection is~ \cite{superbcdr} 
\begin{equation}
    P_{\rm avg}=\frac{P_{\rm eq}}{1+\tau_{\rm tot}/\tau_{\rm b}}+\frac{P_{\rm inj}}{1+\tau_{\rm b}/\tau_{\rm tot}}
\label{eq:Pavg}
\end{equation}
Here $P_{\rm inj}$ is the injected beam polarization, $\tau_{\rm b}$ is the beam lifetime.  For colliding beams at the Z-pole, the beam lifetime is about 2 hours, mainly limited by the radiative Bhabha effect which is correlated to the luminosity. Hence, if the equilibrium polarization is larger than 10\%, $\tau_{tot}$ is longer than 28 hours, still significantly longer than the beam lifetime. In this case, if the polarization of the injected beam is very high (for example, more than 70\%), then the time-averaged beam polarization is approximately the injected beam polarization. In other words, to maintain
a high time-averaged beam polarization, it is essential to realize a high
injected beam polarization, while 
it is only required to achieve a modest equilibrium beam polarization in
the collider ring.

\section{Solenoid-based spin rotators\label{sol} }

According to Eq.~(\ref{eq:spinmotion}), the rotation angle of an electron spin vector in a solenoid can be described by
\begin{equation}
    \varphi_{\rm sol}=\frac{e(1+a)}{m_0c\beta \gamma}\int B_{\parallel}ds
\end{equation}
And the axis of rotation is along the longitudinal direction.
In a horizontal bending dipole, the spin rotation angle is 
\begin{equation}
    \varphi_{\rm dip}=\frac{ea}{m_0 c \beta}\int B_y ds =a\gamma \theta   
\end{equation}
where $\theta$ is the deflection angle of the orbital motion. 
And the axis of the rotation is along the vertical direction.

We use one typical design of solenoid-based spin rotators that consists of a solenoid section and 
a horizontal bending dipole magnet section. In the spin rotator upstream of
the IP (RotatorU),
the solenoid section
rotates the spin vector from the vertical direction
to the radial direction with a rotation angle of $\pi/2$. Then
the horizontal bending magnet section rotates the spin vector by
an odd multiple of $\frac{\pi}{2}$, i.e.  
\begin{equation}
    \varphi=(2K+1)\frac{\pi}{2},~~~K \in Z
    \label{eq:2k+1}
\end{equation}
which can rotates the spin vector from the radial direction to the longitudinal direction at the IP. Rotation of a spin vector from the 
longitudinal
direction at the IP, to the vertical direction in the arc can be realized by
a spin rotator downstream of the 
IP (RotatorD), with a similar setup of RotatorU, but in a reversed
order. Such a pair of spin rotators
can realize longitudinal beam polarization at IPs, while retaining vertical
beam polarization in the arc regions.

\begin{figure}[htbp]
\centering
\subfigure[Anti-symmetric arrangement.]{
\begin{minipage}{1\columnwidth}
\centering
\includegraphics[width=0.9\columnwidth]{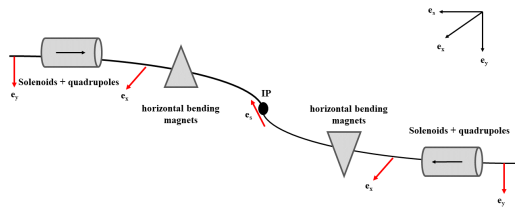} 
\end{minipage}
}
\subfigure[Symmetric arrangement.]{
\begin{minipage}{1\columnwidth}
\centering
\includegraphics[width=0.9\columnwidth]{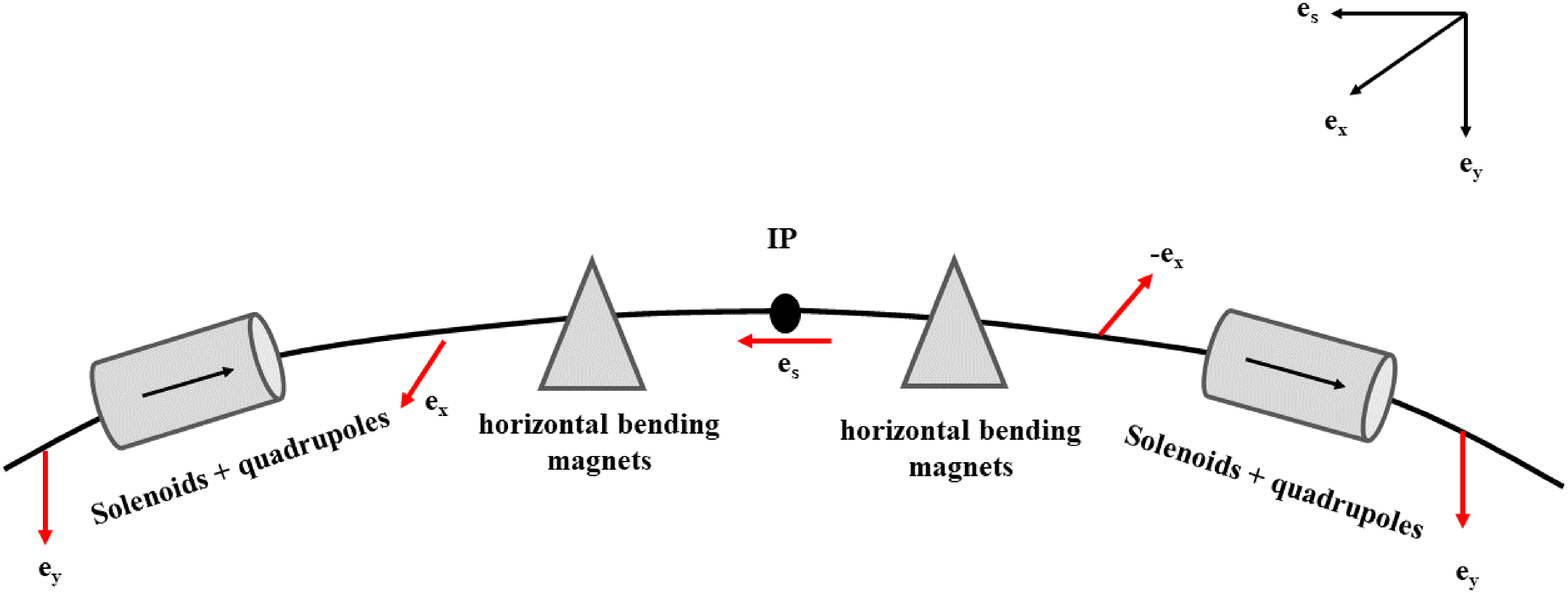}  
\end{minipage}
}
\caption{Two possible arrangements of the spin rotator magnets . The red arrow represents the direction of polarization. Here $\mathbf{e_x}$, $\mathbf{e_s}$, and $\mathbf{e_y}$ are the radial, longitudinal, and vertical unit base vectors, respectively. }
\label{fig:rtt_arrange}
\end{figure}

There are two possibilities of arrangement of a pair of spin rotators
around each IP~\cite{barber},
in terms of whether the spin rotators on both sides of the IP have the same polarity (symmetric arrangement, Fig.~\ref{fig:rtt_arrange}(b)) or opposite polarities (anti-symmetric arrangement, Fig.~\ref{fig:rtt_arrange}(a)),
as illustrated in Fig.~\ref{fig:rtt_arrange}.
Compared with the symmetric arrangement, the anti-symmetric arrangement can exactly rotate the spin vector back to the vertical direction in the arc even when not running at the designed energy. 
In addition,
the anti-symmetric structure also matches the geometry of the CEPC collider ring in the interaction region (IR). Therefore, the anti-symmetric arrangement is preferred when the spin rotators are placed near the interaction region. 

To rotate the spin vector by $\pi/2$ at $E=45.6$ GeV, the 
required integrated field strength of the solenoid magnets is
about $240 {\rm ~T \cdot m}$. For example, if we consider a superconducting solenoid magnetic field of 8~T, the total length of the solenoid magnet is about 30~m. This requires a series of solenoid magnets.

\subsection{Design of the solenoid section}
Solenoids introduce transverse coupling which is detrimental to both the
orbital and spin motion, which must be compensated locally.
There are different compensation schemes
using quadrupoles lens~\cite{decoupleunit1}.
A solenoid compensation unit which consists of two solenoids with 7 normal quadrupoles in between has already been studied in our early work~\cite{xiawh}. However, the compact design of the solenoid compensation unit requires high magnetic field gradient of the decoupled quadrupoles. Reducing the high gradient of quadrupoles would inevitably take up more space. 

\begin{figure}[hbt!]
\centering
%\subfigure[Solenoid compensation unit 1 ~\cite{decoupleunit1}.]
%{
%\includegraphics[width=0.9\columnwidth]{Fig3.eps} 
%\caption{fig1}
%}
%\quad
%\subfigure[Solenoid compensation unit 2 ~\cite{barber}.]
{
\includegraphics[width=1.0\columnwidth]{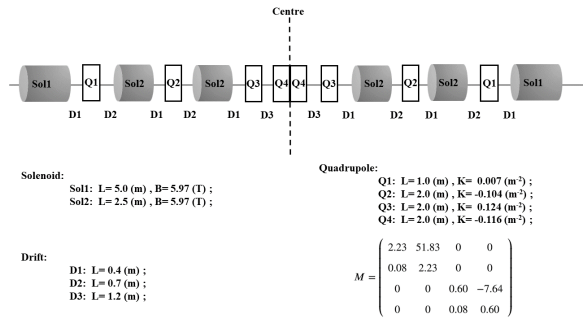}  
}
\caption{ Solenoid compensation unit. Magnet parameters are listed in each subfigure. $\mathbf{M}$ is the transverse transport matrix for each unit. }
\label{fig:decoupleunit}
\end{figure}

In this paper,
we designed another possible solenoid compensation unit (SCU) for the CEPC following the concept proposed in Ref.~\cite{barber}, as shown in Fig.~\ref{fig:decoupleunit}. 
SCU consists of 6 solenoids interleaved with
8 normal quadrupoles, with a spin rotation angle of $\varphi=\frac{\pi}{4}$. The main
parameters of the solenoids and
quadrupoles and the lengths of 
drifts,
as well as the 4-dimensional transfer matrix of a SCU
are also shown in Fig.~\ref{fig:decoupleunit}.

Then, a solenoid module is designed which contains 2 SCUs in
the middle, to rotate the spin vector by $\pi/2$, 
as well as optics matching sections (OM) at both ends, to match with the optical parameters of the CEPC collider ring lattice. Its whole structure is 
\begin{equation}
{\rm OM} + {\rm SCU + D + QF + D + QF + D + SCU } +{\rm OM} 
\nonumber
\end{equation}
Here ``QD'' and ``QF'' mean the defocusing
and focusing quadrupoles, respectively.
``D'' is an abbreviation for drift.
Each OM contains three normal quadrupoles to fit into the optics of the ring lattice.
If the solenoids are turned off, the quadrupoles in the SCU need be retuned to create a normal uncoupled optics with designed phase advances.

\begin{figure}[hbt!]
\centering
%\subfigure[Solenoid module 1.]
%{
%\begin{minipage}{0.5\columnwidth}
%\centering
%\includegraphics[width=0.9\columnwidth]{Fig5.eps} 
%%\caption{fig1}
%\end{minipage}
%}%
%\subfigure[Solenoid module 2.]
{
%\begin{minipage}{1.\columnwidth}
\centering
\includegraphics[width=1.\columnwidth]{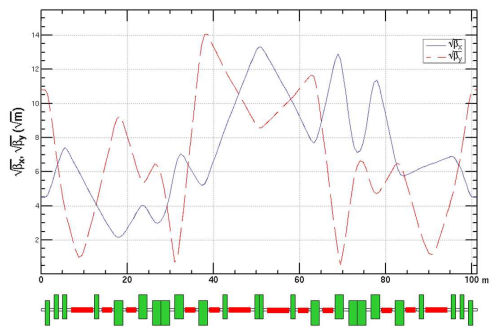}  
%\end{minipage}
}
\caption{Twiss parameters and lattice structures for the solenoid module. Green and red boxes represent normal quadrupoles and solenoid magnets respectively. }
\label{fig:solbeta}
\end{figure}

%The twiss parameters of the regular FODO cell in the arc region of CEPC 
%are $\alpha_x/\alpha_y=0$, $\beta_x/\beta_y=20 ~{\rm m}/100 ~{\rm m}$. 
The OMs and the quadrupoles are adjusted to match the optics at the entrance and exit of solenoid modules. 
Lattice structures and the twiss parameters in the
solenoid module are plotted in Fig.~\ref{fig:solbeta}.

%Table~\ref{tab:solmodule} compares the  key parameters of SM1 and %SM2. SM2 is 220 meters shorter than SM1, and uses much less magnets.

%\begin{table}[!htb]
%\caption{The structure comparison  of the solenoid module 1 and %module 2.}
%\resizebox{\linewidth}{!}
%{\begin{tabular}{@{}lcc@{}} \toprule
%&Solenoid module 1&Solenoid module 2 \\ \bottomrule
%Number of quadrupoles per unit &7&8 \\ 
%Number of solenoids per unit &2&6 \\ 
%Number of solenoid compensation units &10&2 \\ 
%Total length~(m) &320&100 \\ \bottomrule
%\end{tabular} }
%\label{tab:solmodule}
%\end{table}

\section{The insertion scheme\label{opt}}
%In this section, we will optimize the design based on the pros of both insertion schemes. In particular, the solenoid section is placed adjacent to IR. This on one hand eliminates
%the perturbation to the IR optics, on the other hand reduces the depolarization contribution 
%between the IP and the solenoid section.
%This optimization scheme can achieve high level of beam polarization at the IP without affecting the stability of the beam orbital motion. In addition, the solenoid module 2 is utilized here to shorten the length of the spin rotator.

\begin{figure}[hbt!]
\centering
\includegraphics[width=1.\columnwidth]{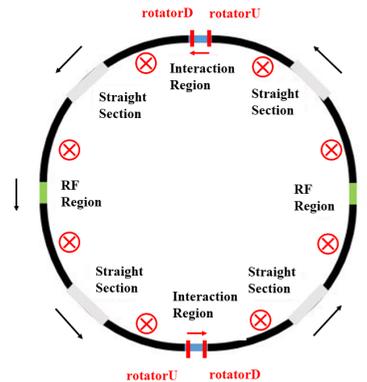} 
\caption{The insertion scheme for spin rotators, taking the positron ring as an illustration. Black arrows represent the direction of beam movement.
 Red arrows represent the direction of $\mathbf{n_0}$. Red lines are spin rotators, while ``rotatorU'' and ``rotatorD'' represent the spin rotators upstream and downstream of the IP, respectively. The black curve represents the arc area.  }
\label{fig:rtt_positions}
\end{figure}

According to Eq.(\ref{eq:2k+1}), the spin rotation around the vertical axis from IP to the solenoid module must be $(2K+1)\frac{\pi}{2}$, $K$ is an integer, to enable the longitudinal polarization at the IP.
This requires adjustments to the layout of the collider ring between the solenoid module insertions. 
We studied different insertion schemes to implement the solenoid module insertions into the lattice.
At the beginning of this investigation, two different insertion schemes were
studied, either inserting the solenoid sections into a short straight section inside the IR~\cite{xiawh}, or placing the solenoid sections in the adjacent
long straight sections in the ``symmetric arrangement'', both schemes however yield unsatisfactory results. The first scheme
introduces strong perturbation to the 
layout of the IR and the optics performance, though the depolarization effect is moderate.
The second scheme introduces less perturbation to the optics,
but suffers from strong depolarization, because the polarization direction rotates many turns by the horizontal bending magnets between
the IP and the solenoid section, which enhances the spin diffusion effects.
Finally, based on the pros of both insertion schemes, an optimized insertion scheme was found. 
The solenoid sections are placed adjacent to the IR in Fig.~\ref{fig:rtt_positions}, forming an anti-symmetric arrangement.
This on one hand eliminates
the perturbation to the IR optics, on the other hand reduces the depolarization contribution 
between the IP and the solenoid section.
This optimization scheme can achieve high level of beam polarization at the IP without affecting the stability of the beam orbital motion.

\begin{figure}[hbt!]
\centering
\includegraphics[width=1.\columnwidth]{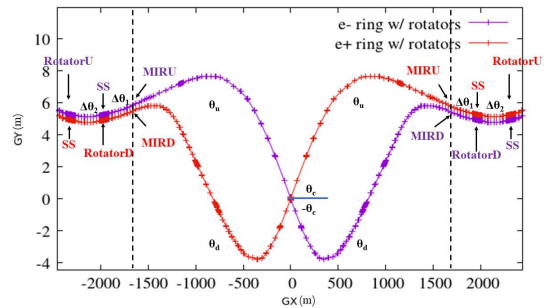} 
\caption{Geometry of the $e^+/e^-$ rings with spin rotators of the insertion scheme around the IR. }
\label{fig:geo_IR4}
\end{figure}

As shown in Fig.~\ref{fig:rtt_positions}, a pair of spin rotators are implemented around each IP, which enables longitudinal polarization of the beam at both IPs. The polarity setting of the solenoid magnetic field in the spin rotators ensures that the  directions of $\mathbf{n_0}$ in the two arc regions on the left and right of the IP are the same, and $\nu_0 \approx a\gamma_0$ still holds, which does not affect the beam energy measurement based on resonant depolarization.
Fig.~\ref{fig:rtt_positions} takes the positron ring as an example. The spin rotator settings in the electron ring and the positron ring are just symmetrical around the IP as plotted in Fig.~\ref{fig:geo_IR4}. The helicity of the colliding polarized beams  could be adjusted by varying the helicity of the injected polarized beams from the polarized particle sources, which will be studied  separately.

 The layout and the optics of IR will remain unchanged. Just upstream of the IR, we insert a bending angle compensation section $\Delta\theta_1$, a straight section (SS) of the same length as the solenoid module, a bending angle compensation section $\Delta\theta_2$ and the
solenoid module (rotatorU) in sequence. Just downstream of the IR, we insert a bending angle compensation section $\Delta\theta_1$, the solenoid module (rotatorD), a bending angle compensation section $\Delta\theta_2$ and
a SS in sequence. More details of the optimized insertion scheme are plotted in Fig.~\ref{fig:geo_IR4}.

In Fig.~\ref{fig:geo_IR4}, the default parameters in the interaction region (IR) of the CEPC lattice are
\begin{itemize}
    \item $\theta_c=16.50$~mrad, half of the crossing angle between $e^+$ ring and $e^-$ ring at the IP;
    \item $\theta_u=-19.21$~mrad, the total horizontal bending angle from the IP to the end of the interaction region ``MIRU"; 
    \item $\theta_d=13.79$~mrad, the total horizontal bending angle from the IP to the end of the interaction region ``MIRD";
\end{itemize}

Then the longitudinal polarization condition requires
\begin{eqnarray}
 a\gamma_0(\theta_u+\Delta\theta_1+\Delta\theta_2)&=&-\frac{\pi}{2} \nonumber \\
 a\gamma_0(\theta_d+\Delta\theta_1)&=&\frac{\pi}{2}. 
\end{eqnarray}
We can get $\Delta\theta_1=1.39$~mrad, $\Delta\theta_2=2.65$~mrad when $a\gamma_0=103.5$.
 
The lattice structure of SS is basically the same as that of the solenoid module. However, the solenoid magnets are replaced with drifts of same length. And the quadrupoles inside the SS are retuned to create a normal uncoupled optics which is matched to the collider ring.

The angle compensation sections $\Delta\theta_1$ and $\Delta\theta_2$ are designed as achromat modules. Each section consists of 4 identical FODO cells with $\frac{\pi}{2}$ phase advances for both planes, to ensure that the dispersion function inside the solenoid module is zero. The dipoles in these FODO cells are as long as the dipoles in the arc FODO cells.
However, the magnetic field strength is about an order of magnitude weaker than the arc dipoles.

We then retuned several quadrupoles to recover the fractional part of the betatron tunes of the lattice. And the deflection angles of the dipoles in the rest part of the collider ring are reduced accordingly to ensure the total horizontal deflection angle of the entire ring is $2\pi$. Next, we evaluate the performance of this lattice design via detailed simulations.

\subsection{Polarization simulation by Bmad/PTC}

Spin rotators can drive betatron sideband spin resonances and lead to a lower equilibrium polarization in the nearby region even for a perfect storage ring. The electrons and positrons execute betatron oscillations, and they are subjected to additional magnet fields as they go through quadrupoles and solenoids. And their spin precession acquires a small additional orbit-dependent component. In particular, between the solenoid sections around each IP,
$\mathbf{n_0}$ is in the horizontal plane, the horizontal betatron oscillations in quadrupole magnets, lead to spin precession around the vertical direction,
perpendicular to $\mathbf{n_0}$. Hence, this drives first-order betatron sideband spin resonances $\nu_0=k\pm\nu_x$.

In a realistic storage ring, the solenoid magnetic field may be not perfectly compensated due to magnet errors. These magnet errors also drive spin resonances 
and lead to a reduced equilibrium beam polarization.
Assuming that these magnet errors follow a Gaussian distribution in the solenoid compensation unit, we introduced relative field errors for solenoids and quadrupoles with a root-mean-squared value of 0.05\%, and relative roll errors for quadrupoles with a root-mean-squared value of 0.01\%. 

\begin{figure}[hbt!]
\centering
\includegraphics[width=1.\columnwidth]{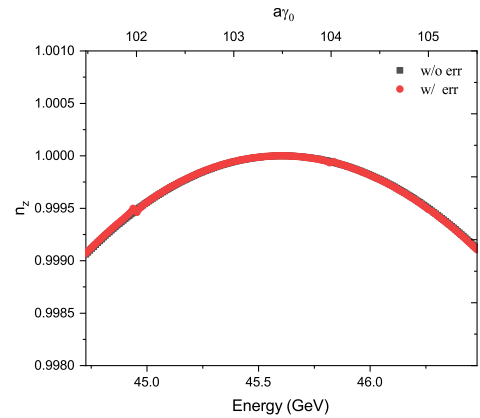}
\caption{The longitudinal projection of $\mathbf{n_0}$ at the IP for different beam energies, for the lattices with and without magnet errors.}
\label{fig:n0}
\end{figure}

\begin{figure}[hbt!]
\centering
\includegraphics[width=1.\columnwidth]{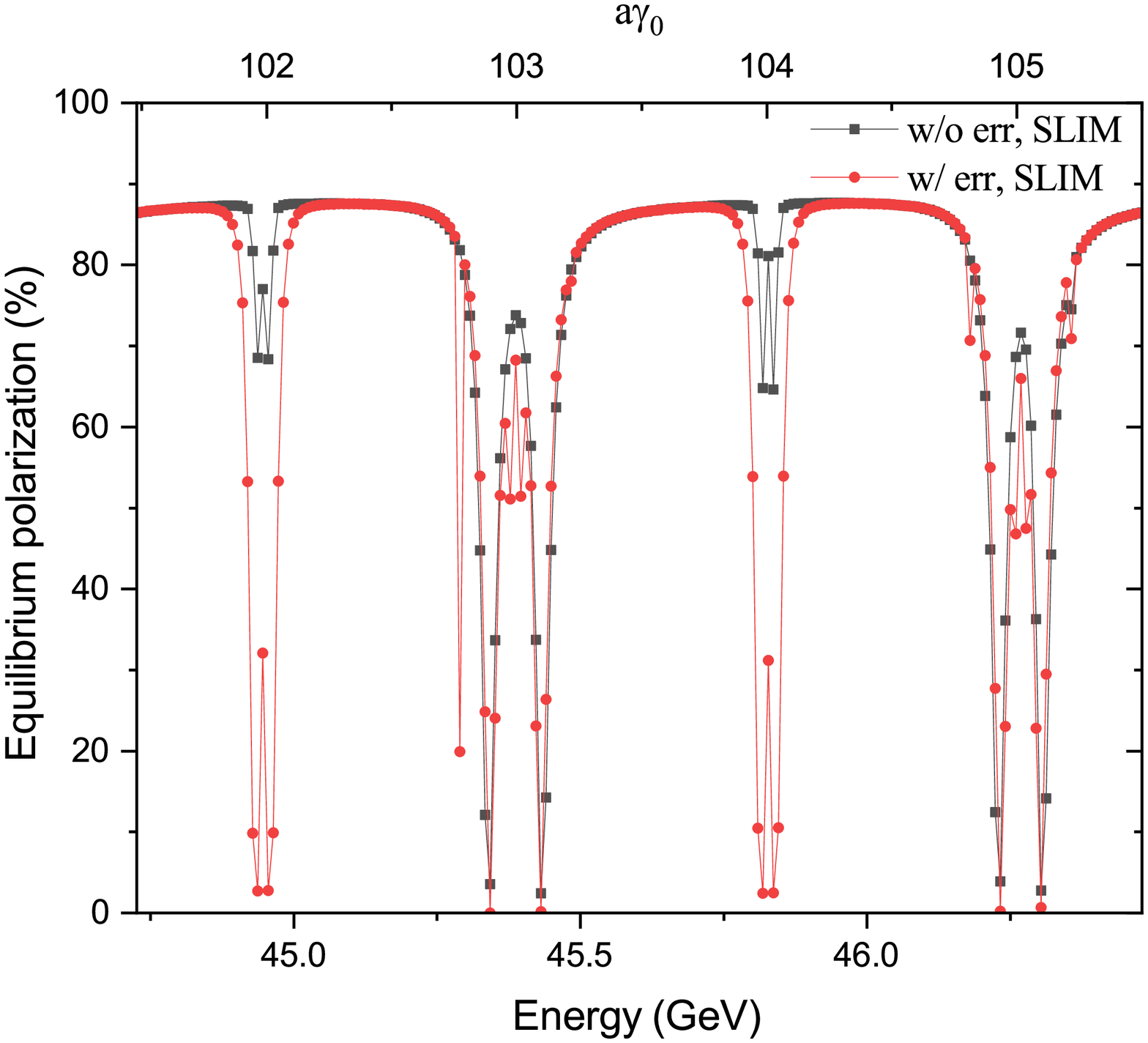}
\caption{The equilibrium beam polarization at different beam energies, for the lattices with and without magnet errors, simulated using SLIM in Bmad.}
\label{fig:slim}
\end{figure}

\begin{figure}[hbt!]
\centering
\includegraphics[width=1.\columnwidth]{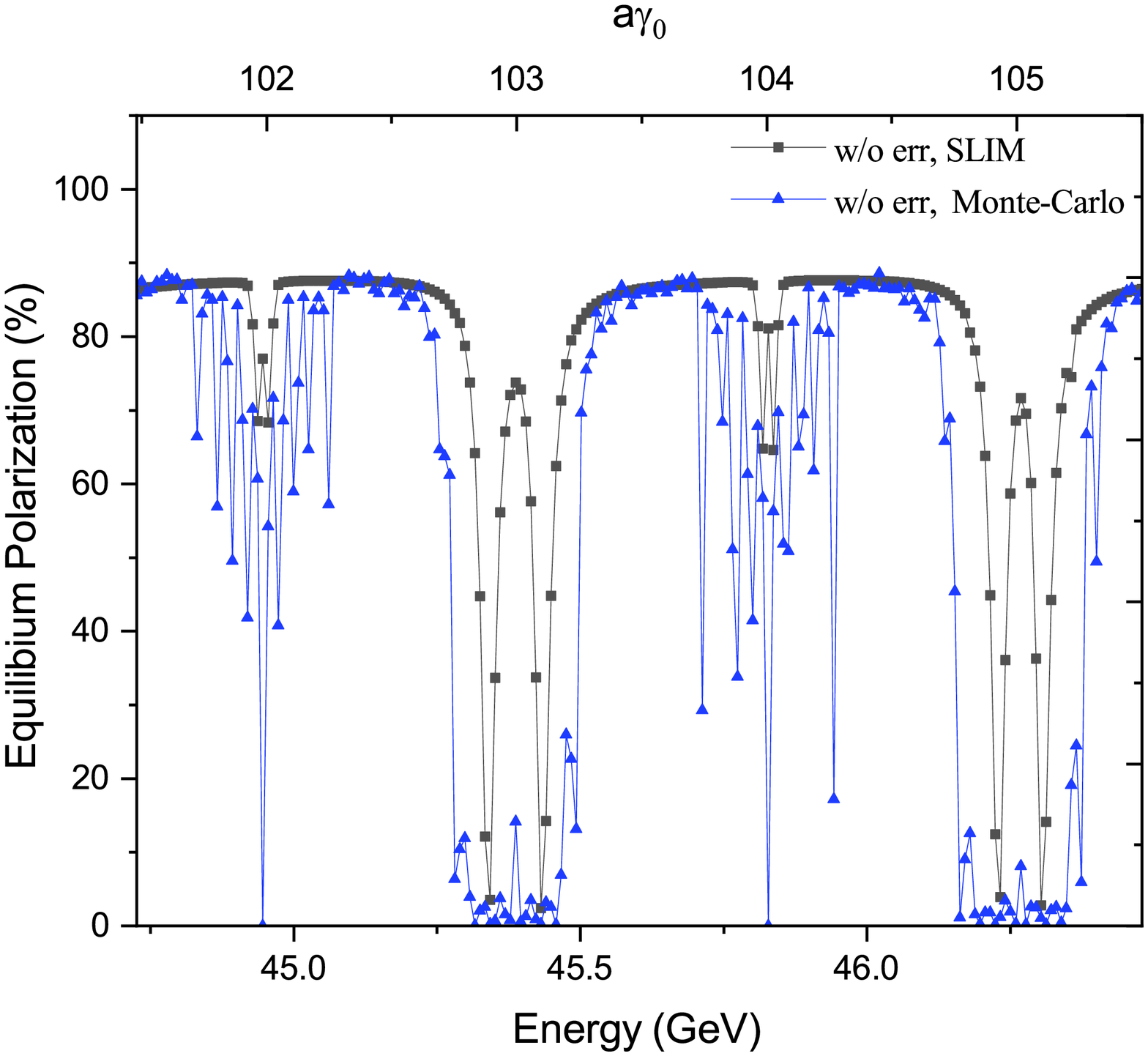}
\caption{The equilibrium beam polarization for the lattice without magnet errors, simulated using SLIM in Bmad and Monte-Carlo method in PTC, respectively. The step size $\Delta a\gamma_0=0.02$.}
\label{fig:MC}
\end{figure}

We compared the longitudinal projection of beam polarization at the IP and the equilibrium polarization at the Z-pole. 
As shown in Fig.~\ref{fig:n0}, 
the longitudinal projection of $\mathbf{n_0}$ at the IP is symmetrically distributed on both sides of the designed energy, $a\gamma_0=103.5$.
The influence of the magnet errors is negligible.
 Fig.~\ref{fig:slim} shows the comparison between the equilibrium beam polarization simulated with SLIM inside Bmad, for the cases without
 magnet errors and with magnet errors.
The equilibrium polarization degree can be very high when the beam energy is far from spin resonances.  The magnet errors enhance the depolarization  near spin resonances. However, they have negligible effect on the degree of polarization when the fractional part of $a\gamma_0$ is near 0.5. This shows the robustness of the design against machine imperfections.
Besides, Fig.~\ref{fig:MC} shows the influence of higher-order spin resonances on equilibrium beam polarization, relative to the simulation results with SLIM. Here, depolarization effects of the higher-order spin resonances are simulated by a Monte-Carlo method based on PTC~\cite{DUANZ}. The
spin resonance regions where the equilibrium polarization level is low becomes wider and higher-order synchrotron sideband spin resonances are visible.

Other kinds of machine imperfections also affect the orbital motion and the degree of equilibrium polarization, a systematic approach of error correction
is under development~\cite{wangbi}.
Evaluation of the influence of these errors on the equilibrium polarization for a CEPC lattice without spin rotators is reported in~\cite{tobepub},
which shows a relatively high beam polarization is achievable at the Z-pole energy after the error correction.
In future studies, we will introduce machine imperfections for the whole collider ring with spin rotators and evaluate the performance after a dedicated
correction scheme.

The time-averaged beam polarization is determined by the equilibrium polarization and the injected beam polarization according to Eq.(\ref{eq:Pavg}), which can also be expressed as
\begin{equation}
    P_{\rm avg}=\frac{P_{\rm eq}}{1+\frac{P_{\rm eq}}{P_{\rm \infty}}\frac{\tau_{\rm p}}{\tau_{\rm b}}}+\frac{P_{\rm inj}}{1+\frac{P_{\rm \infty}}{P_{\rm eq}}\frac{\tau_{\rm b}}{\tau_{\rm p}}}.
\label{eq:Pavg1}
\end{equation}
For the CEPC at the Z-pole, $\tau_{\rm p}\approx 260$ hours and $\tau_{\rm b}\approx 2$ hours. If the injected beam polarization degree is
70\%, the equilibrium polarization only needs to be greater than 1.8\% to maintain the time-averaged polarization larger than 50\%. There is quite a large
margin from the current simulation results of the equilibrium beam polarization, which reserves tolerance for other potential depolarization mechanisms, for 
example the influence from beam-beam interaction~\cite{kondratenko}, which will be evaluated separately.
The above simulations indeed confirm the analysis that depolarization effects could be largely suppressed in such an anti-symmetric structure in Ref.~\cite{nikitin2,nikitin3}.

\subsection{Influence to orbital motion}
\begin{figure}[hbt!]
\centering
\subfigure[Near the rotatorD.]
{
\includegraphics[width=\columnwidth]{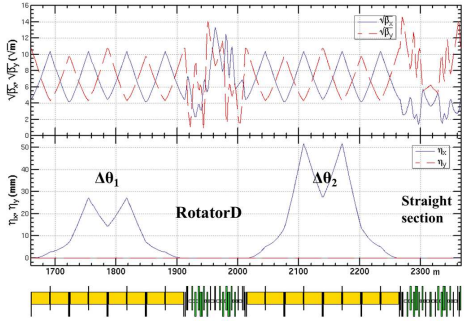}
}

\subfigure[Near the rotatorU.]
{
\includegraphics[width=\columnwidth]{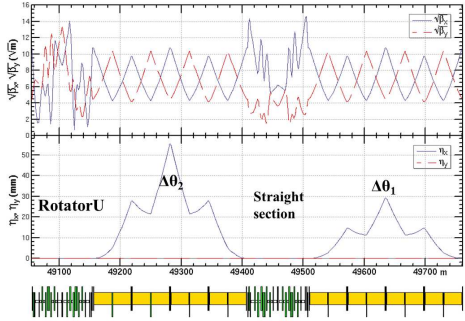}
}
\caption{The optics of the lattice region near spin rotator insertions.}
\label{fig:opticsrtt4}
\end{figure}

\begin{figure}[hbt!]
\centering
\subfigure[Collider ring lattice w/o spin rotators]
{
\includegraphics[width=\columnwidth]{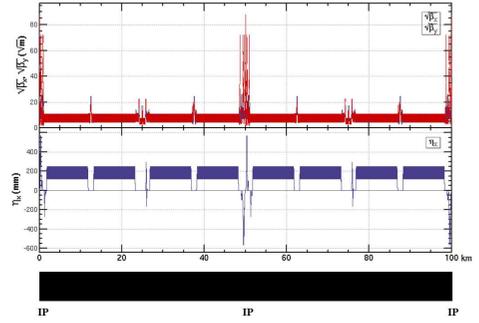}
}

\subfigure[Collider ring lattice with spin rotators.]
{
\includegraphics[width=\columnwidth]{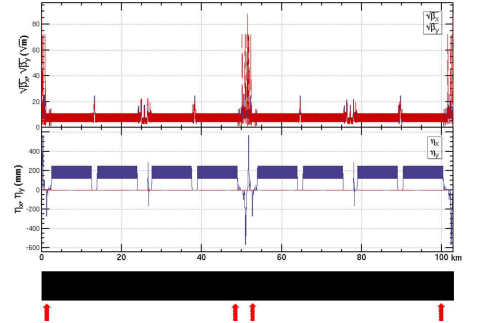}
}
\caption{The optics of the collider ring without and with spin rotators. The red arrows  mark the position of spin rotators.}
\label{fig:opticsrtt4_1}
\end{figure}

The optics of the spin rotator insertions rotatorD and rotatorU are shown in Fig.~\ref{fig:opticsrtt4}. 
The optics of the collider ring before and after insertion of the spin rotators are shown in Fig.~\ref{fig:opticsrtt4_1}. The betatron motion is only locally coupled in the spin rotators, while the optics of the rest of the collider ring remains undisturbed.

To evaluate the influence of the spin rotators on the orbital motion performance.
We consider three
different cases. The first is the CEPC CDR lattice without spin rotators (denoted
as CDR lattice), the second is the CEPC CDR lattice with spin rotator insertions
to realize longitudinal beam polarization (denoted as Solenoid On), the third is
the CEPC CDR lattice with spin rotator insertions, but the solenoids are turned off and 
quadrupoles are retuned to recover the lattice (denoted as Solenoid Off).

Table \ref{tab:cdrvsrtt2} compares the key orbital parameters between these
three cases. There is an increase in the integer parts of the betatron tunes of the latter
two cases, and some tiny differences in
the momentum compact factor and $U_0$.
The circumference of the collider ring increases by approximately 2.8 kilometers in the
presence of the spin rotators. There is no difference in the orbital parameters between Solenoid off and Solenoid on.
\begin{table*}[!htb]
\caption{The comparison of several key orbital parameters  between the insertion scheme and the CDR lattice at the Z-pole .}
\resizebox{\linewidth}{!}
{\begin{tabular}{@{}lccc@{}} \toprule
&CDR Lattice& Solenoids On & Solenoids Off\\ \hline
Tunes $\nu_x/\nu_y/\nu_z $&363.11/365.22/0.028 &381.11/383.22/0.028  &381.11/383.22/0.028 \\
%$\epsilon_x/\epsilon_y/\epsilon_z$&0.18~nm/0/$8.86\times10^{-7}$~m&0.176~nm/0/$8.86378\times10^{-7}$~m&0.176~nm/0/$8.86181\times10^{-7}$~m \\
Emittances $ \epsilon_x/\epsilon_z$&0.18~nm/$0.886~\mu$m&0.18~nm/$0.886~\mu$m&0.18~nm/$0.886~\mu$m \\
%Momentum compact $\alpha_p$&$1.11\times 10^{-5}$&$1.0740\times 10^{-5}$&$1.0740\times 10^{-5}$ \\
Momentum compact factor $\alpha_p$&$1.11\times 10^{-5}$&$1.07\times 10^{-5}$&$1.07\times 10^{-5}$ \\
%First order chromaticity $\xi_x/\xi_y$& 0/0& &\\
Circumference~(m) &100016.35&102841.95&102841.95 \\
SR energy loss per turn $ U_0$~(MeV)&35.47&35.91&35.91 \\
$\beta$-function at IPs~$\beta_x^{\star}/\beta_y^{\star}$&0.2/0.001&0.2/0.001&0.2/0.001 \\ \bottomrule
\end{tabular} }
\label{tab:cdrvsrtt2}
\end{table*}

\begin{figure}[hbt!]
\centering
\subfigure[DA in the horizontal direction. ]
{
\includegraphics[width=\columnwidth]{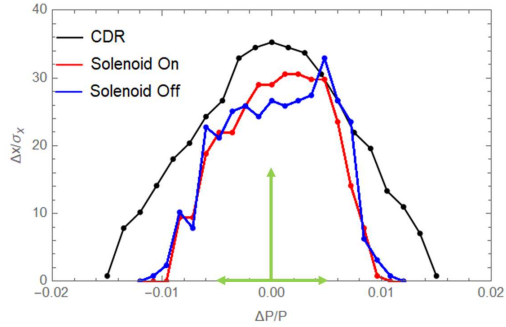}
}
\quad
\subfigure[DA in the vertical direction.]
{
\includegraphics[width=\columnwidth]{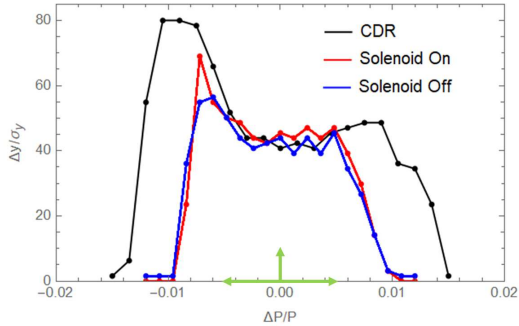}
}
\caption{Comparison of the dynamic apertures between the CDR lattice and the CDR lattice with spin rotators (solenoid on/off). The  green arrows represent the requirement on dynamic apertures at the Z-pole. }
\label{fig:DA}
\end{figure}

Furthermore, we examine the effect of the spin rotator on the nonlinear
performance of the lattice. 
%The first order chromaticities are corrected by adjusting the sextupoles in the arc sections, but without further nonlinear optimization.
The leading orders of the chromaticity have been corrected by adjusting the sextupoles in the arc sections with SAD.
We tracked the particles for one transverse damping time (about 2600 turns) to
obtain the dynamic apertures, are shown in Fig.~\ref{fig:DA}. Spin rotators in the collider ring leads to a moderate shrink of the dynamic apertures. Since the requirement on dynamic aperture at the Z-pole is $17\sigma_x \times 9 \sigma_y \times 0.49\%$ according to the CEPC CDR, the shrink is still acceptable. The change in the dynamic aperture  for Solenoid Off is tiny  compared to that for Solenoid On. This means that the effect of the solenoid magnets on the nonlinear
performance of the lattice is compensated by the quadrupoles in the solenoid compensation unit.
The dynamic aperture can be further optimized using more families of sextupoles, which is a conventional step in the lattice optimization.

According to the CEPC CDR, the CEPC will work at three different energy zones.
In this paper, the spin rotator presented is specifically optimized for the Z energy, but not capable
to realize longitudinal polarization at the W and Higgs energies.
The possibility of a spin rotator design
covering a larger energy range (for example covering the W energy and even the Higgs energy as well)
is beyond the scope of this work.
Here, we tentatively assume 
the solenoid magnets in the spin rotators will be turned off when working at the W
and Higgs energy zones. Then, the quadrupoles in SCUs
need to be retuned to create a normal uncoupled optics.
Note that the same machine layout is maintained for different beam energies,
while a fine adjustment of the optics is needed, 
for example in the IR region to achieve different $\beta^{\star}$ at IPs.
Based on the above considerations, we also studied the influence of these spin rotators
for the W and Higgs energy zones. 

\begin{figure}[hbt!]
\centering
\subfigure[DA in the horizontal direction. ]
{
\includegraphics[width=\columnwidth]{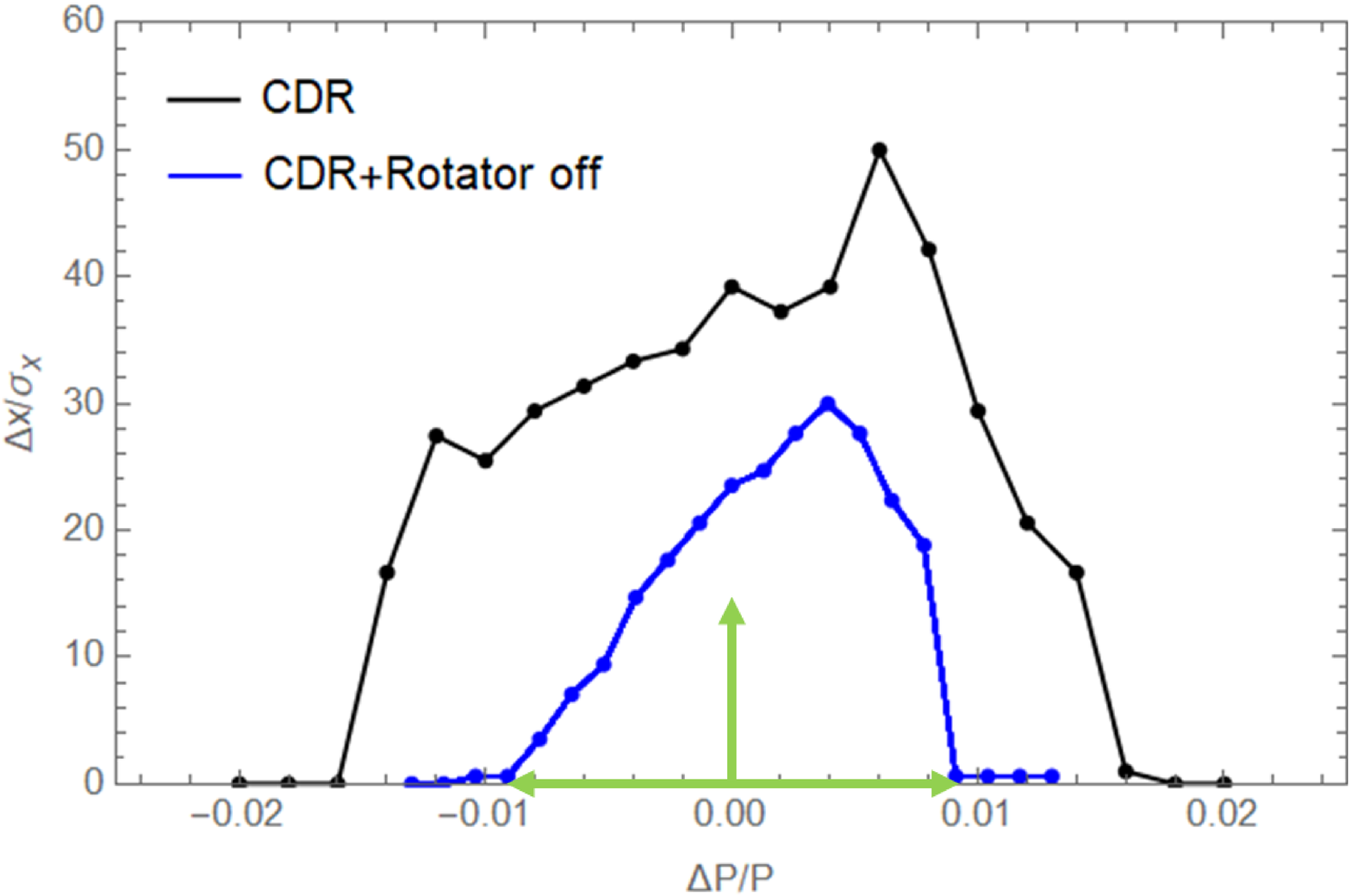}
}
\quad
\subfigure[DA in the vertical direction.]
{
\includegraphics[width=\columnwidth]{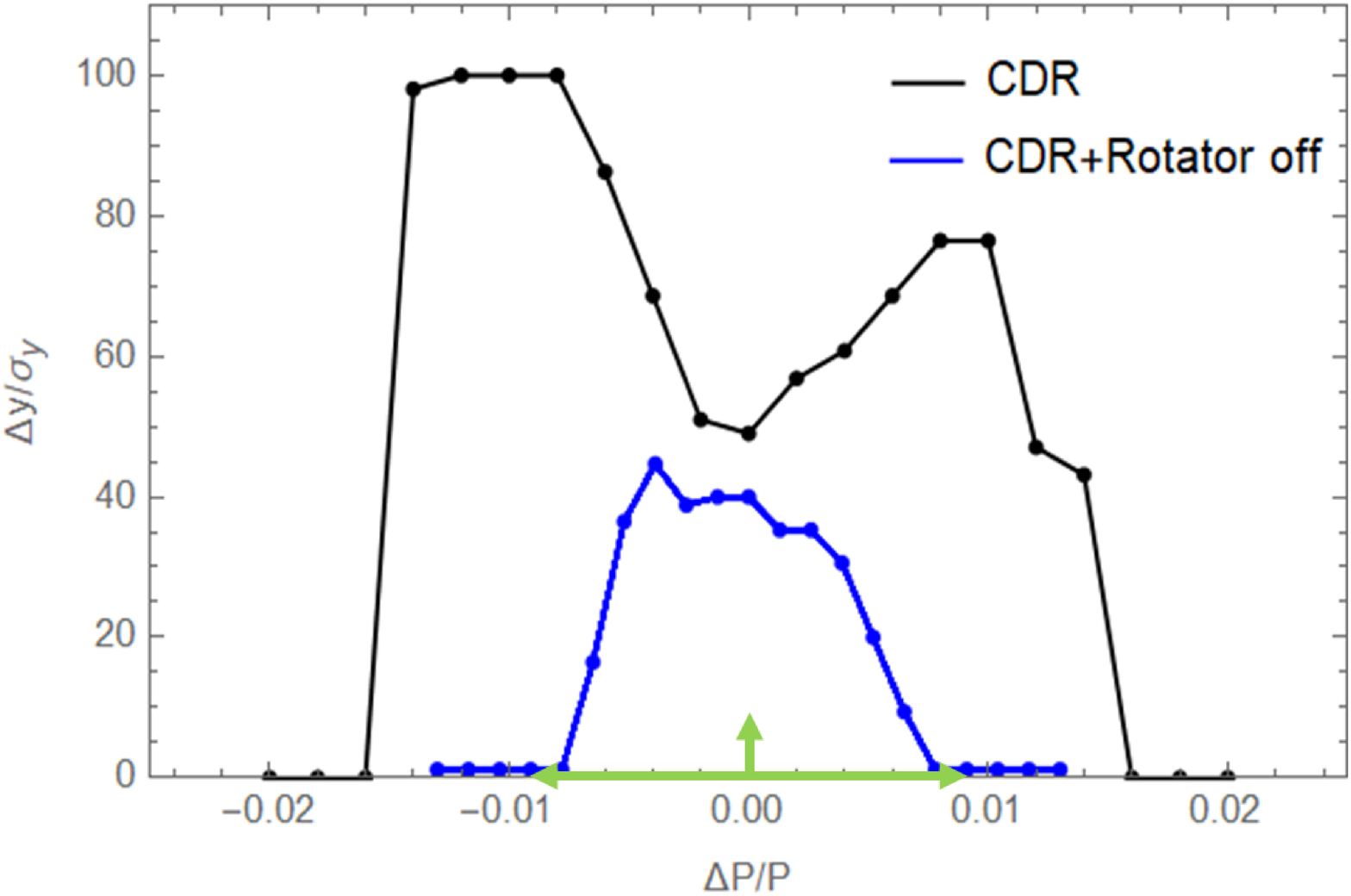}
}
\caption{Comparison of the dynamic apertures between the CDR lattice and the CDR lattice with spin rotators for the CEPC at the W energy zone.  The requirement on the dynamic aperture is $15\sigma_x \times 9 \sigma_y \times 0.9\%$. }
\label{fig:DAW}
\end{figure}

\begin{figure}[hbt!]
\centering
\subfigure[DA in the horizontal direction. ]
{
\includegraphics[width=\columnwidth]{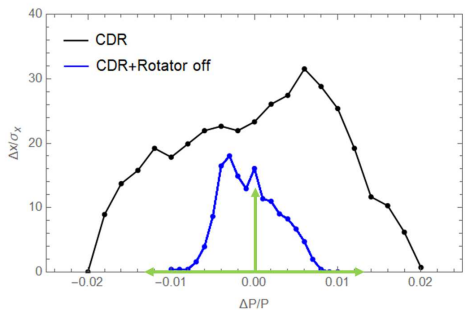}
}
\quad
\subfigure[DA in the vertical direction.]
{
\includegraphics[width=\columnwidth]{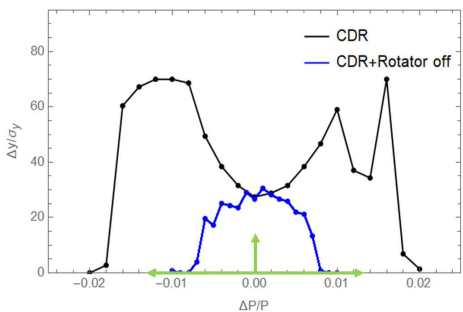}
}
\caption{Comparison of the dynamic apertures between the CDR lattice and the CDR lattice with spin rotators for the CEPC at the Higgs energy zone.  The requirement on the dynamic aperture is $13\sigma_x \times 15 \sigma_y \times 1.35\%$. }
\label{fig:DAH}
\end{figure}

Table~\ref{tab:rtt_w_higgs} shows the comparison of several key orbital parameters between the CDR lattice and the CDR lattice with rotators at W and Higgs energy zone. And the differences of those parameters are acceptable.
The influences of these rotators on the dynamic apertures for W and Higgs energy zones are shown in Fig.~\ref{fig:DAW} and Fig.~\ref{fig:DAH}, respectively. Here, the leading orders of the chromaticity have been corrected by adjusting the sextupoles in the arc sections with SAD.
Spin rotators in the collider ring leads to a shrink of the dynamic apertures for W and Higgs,
in particular the momentum acceptances are smaller than the requirements.
Hence, more comprehensive optimization of the dynamic apertures is foreseen, using more faimilies of
sextupoles.  

\begin{table*}[!htb]
\caption{ Comparison of several key orbital parameters between the CDR lattice and the CDR lattice with rotators at W and Higgs energy zone.}
\centering
\resizebox{\linewidth}{!}
{\begin{tabular}{@{}lcccc@{}} \toprule
& \multicolumn{2}{c}{W}& \multicolumn{2}{c}{Higgs}\\ \hline
&CDR&CDR+Rotators&CDR&CDR+Rotators \\ \hline
Tunes~$\nu_x/\nu_y/\nu_z $& 363.11/365.22/0.039&379.11/381.22/0.039& 363.11/365.22/0.065&379.11/381.22/0.065\\
Emittances~$\epsilon_x/\epsilon_z$&0.54~nm/1.94~$\mu$m&0.53~nm/1.95~$\mu$m& 1.21~nm/2.66~$\mu$m &1.21~nm/2.65~$\mu$m \\
Momentum compact factor $\alpha_p$&$1.11\times 10^{-5}$&$1.07\times 10^{-5}$ &$1.11\times 10^{-5}$&$1.07\times 10^{-5}$\\
Circumference~(m) &100016.35&102841.95&100016.35&102841.95 \\
SR energy loss per turn $ U_0$~(GeV)&0.34&0.34 &1.73&1.72\\ 
$\beta$-function at IPs~$\beta_x^{\star}/\beta_y^{\star}$&0.36/0.0015&0.36/0.0015&0.36/0.0015&0.36/0.0015\\ \bottomrule
\end{tabular} }
\label{tab:rtt_w_higgs}
\end{table*}

\section{Potential optimization of the spin rotators insertion \label{further}}
In the above study, we present a complete spin rotator design and its insertion scheme. However the insertion scheme still has the potential for further optimization, especially in terms of the total length.

As we described earlier, the dipole length in the bending angle compensation sections is the same as the arc dipole, while the magnetic field is an order of magnitude weaker. Hence, we can reduce the length of the bending angle compensation sections. For example, if we reduce the length of this sections by a factor of five, from 250 meters to 50 meters. Then the beta functions at the beginning and end of the sections will also be reduced by five times. We can insert triplet cells at both ends of the section to  match the optical parameters to the rest of the collider ring. According to a simple calculation, each triplet is about 40 meters long with regular quadrupoles. The length of one bending angle compensation section can be reduced from 250 meters to 130 meters. Therefore, the increase of the lattice circumference can be reduced to about 1.8 kilometers.

\begin{figure}[hbt!]
\centering
\includegraphics[width=1.\columnwidth]{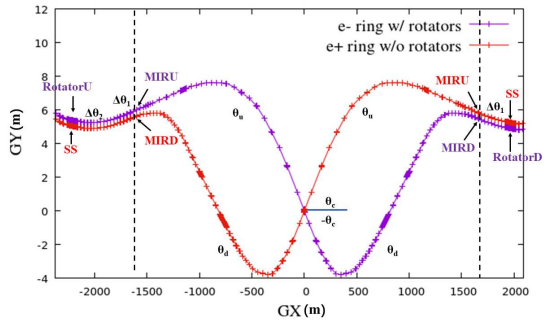} 
\caption{Geometry of the $e^+/e^-$ rings with spin rotators only for $e^-$ beams. }
\label{fig:geo_IR4_e-rtt}
\end{figure}

Based on the assumption that both electrons and positrons can be generated by polarized particle source, we design spin rotators for $e^+$ and $e^-$ rings for the CEPC. However, it is also beneficial for physics programs if longitudinal polarization is only achieved for electrons, if matured polarized $e^+$ source is still not available. In this case, spin rotators are not needed in the $e^+$ ring. Fig.~\ref{fig:geo_IR4_e-rtt} shows the geometry of the $e^+/e^-$ rings with spin rotators only in the $e^-$ ring.
Upstream of the IR of the $e^-$ ring,  the bending angle compensation section $\Delta\theta_1$ and $\Delta\theta_2$ are needed following the spin rotator (rotatorU). Downstream of the IR, only the bending angle compensation section $\Delta\theta_1$ is needed prior to the spin rotator (rotatorD). 
Considering the length reduction potential of the bending angle compensation sections, the increase of the circumference can be further reduced to 1.18 km.

In addition, the bending angle compensation section $\Delta\theta_1$ can be incorporated into the IR, in case of a redesign of the CEPC lattice.
Thereby, the increase of the circumference can be further reduced to 660~m for longitudinal polarized $e^-$ beam alone or 1.28 km for longitudinal polarized $e^-/e^+$  beams.

If the crossing angle at the IP can be changed on the assumption that the CEPC collider ring lattice will be redesigned for running with longitudinal polarization in the future. The following formulas need to be satisfied
\begin{eqnarray}
 a\gamma_0(\theta_u+\Delta\theta)&=&-\frac{\pi}{2} \nonumber \\
 a\gamma_0(C_d\theta_d+\Delta\theta)&=&\frac{\pi}{2} \nonumber\\
 \theta''_{{\rm IR}}=-\theta'_c+C_d\theta_d +\Delta\theta&=&\theta'_c+ \theta_u+ \Delta\theta
\end{eqnarray}
Here, $\theta''_{{\rm IR}}$ is the angle between the end direction of the IR and the horizontal coordinates of global geometry.
All the bending angles of the dipoles in the downstream of the IR need to be multiplied  by $C_d$, which equals to 0.808. And the crossing angle is $2\theta'_c=30.35$~mrad. Here the bending angle compensation section $\Delta\theta=4.04$~mrad, which can be incorporated into the IR. In this case, the increase of the collider ring length is only contributed by the four spin rotators. Hence, the increase can be reduced to about 400~m for longitudinal polarized $e^-$ beam alone or for longitudinal polarized $e^-/e^+$  beams.

\section{Summary}
As a key research topic in implementing longitudinal polarized beams for the CEPC at the Z-pole, the designs of solenoid-based spin rotators for the CEPC collider rings are discussed in this article. We presented an optimized insertion schemes with a complete spin rotator design
in the CEPC CDR lattice. Simulation results shows that high longitudinal beam polarization can be achieved at the IP and the influence to the orbital motion is acceptable.

At this stage our models do not have misalignments for the whole collider ring. And the detector solenoids are not included either. However, the simulation of realistic misalignments, the modelling of the correction of the orbital imperfections and the depolarizing effects of the detector solenoids are very important for beam polarization at ultra high beam energies. These topics will be studied in the future on the basis of this simulation framework, and will be an integral part of investigation of feasibility of attaining high beam polarization at the CEPC.

There is still room for further optimization of the dynamic apertures and the length of spin rotators. 
As we discussed in Chapter~\ref{further}, the length growth of the collider ring can be reduced to 1.8~km for $e^+/e^-$ longitudinal polarization and to 1.18~km
for $e^-$ longitudinal polarization based on the current CEPC CDR lattice design. In case of a redesign of the CEPC lattice, if the crossing angle can be 
modified to 30.35~mrad and the IR region are adjusted accordingly, the total length occupied by the insertion of the spin rotator can be reduced to
about 400~m for longitudinal polarized $e^-$ beam alone or for longitudinal polarized $e^-/e^+$ beams.

\backmatter

\bmhead{Acknowledgments}
The authors are especially grateful to Dr. Sergei Nikitin for the helpful discussion and the proofreading.
This study was supported by National Natural Science Foundation of China (Grant No. 11975252); National Key Program for S\&T Research
and Development (Grant No. 2016YFA0400400 and 2018YFA0404300); Key Research Program of Frontier Sciences, CAS (Grant No. QYZDJ-SSW-SLH004); Youth Innovation Promotion Association CAS (No. 2021012).

\section*{Declarations}
On behalf of all authors, the corresponding author states that there is no conflict of interest.
\noindent

\bibliography{sn-bibliography}% common bib file


\end{document}